\documentclass[a4paper,twoside, titlepage]{article}
\usepackage[utf8]{inputenc}
\usepackage[T1]{fontenc}
\usepackage{subcaption}
\usepackage{calc}
\usepackage{amssymb}
\usepackage{amstext}
\usepackage{amsmath}
\usepackage{amsthm}
\usepackage{multicol}
\usepackage{pslatex}
\usepackage{apalike}
\usepackage{graphicx}

\usepackage{multirow} 

\makeatletter
\def\@citex[#1]#2{%
	\let\@citea\@empty
	\@cite{\@for\@citeb:=#2\do
		{\@citea\def\@citea{;\penalty\@m\ }%
			\edef\@citeb{\expandafter\@firstofone\@citeb}%
			\if@filesw\immediate\write\@auxout{\string\citation{\@citeb}}\fi
			\@ifundefined{b@\@citeb}{\mbox{\reset@font\bfseries ?}%
				\G@refundefinedtrue
				\@latex@warning
				{Citation `\@citeb' on page \thepage \space undefined}}%
			{\csname b@\@citeb\endcsname}}}{#1}}
\makeatother

\usepackage{SCITEPRESSHAL}

\title{A Formally Correct and Algorithmically Efficient \\LULC change Model-Building Environment}

\author{\authorname{Fran\c cois-R\'emi Mazy\sup{1}\thanks{\includegraphics[width=0.3cm]{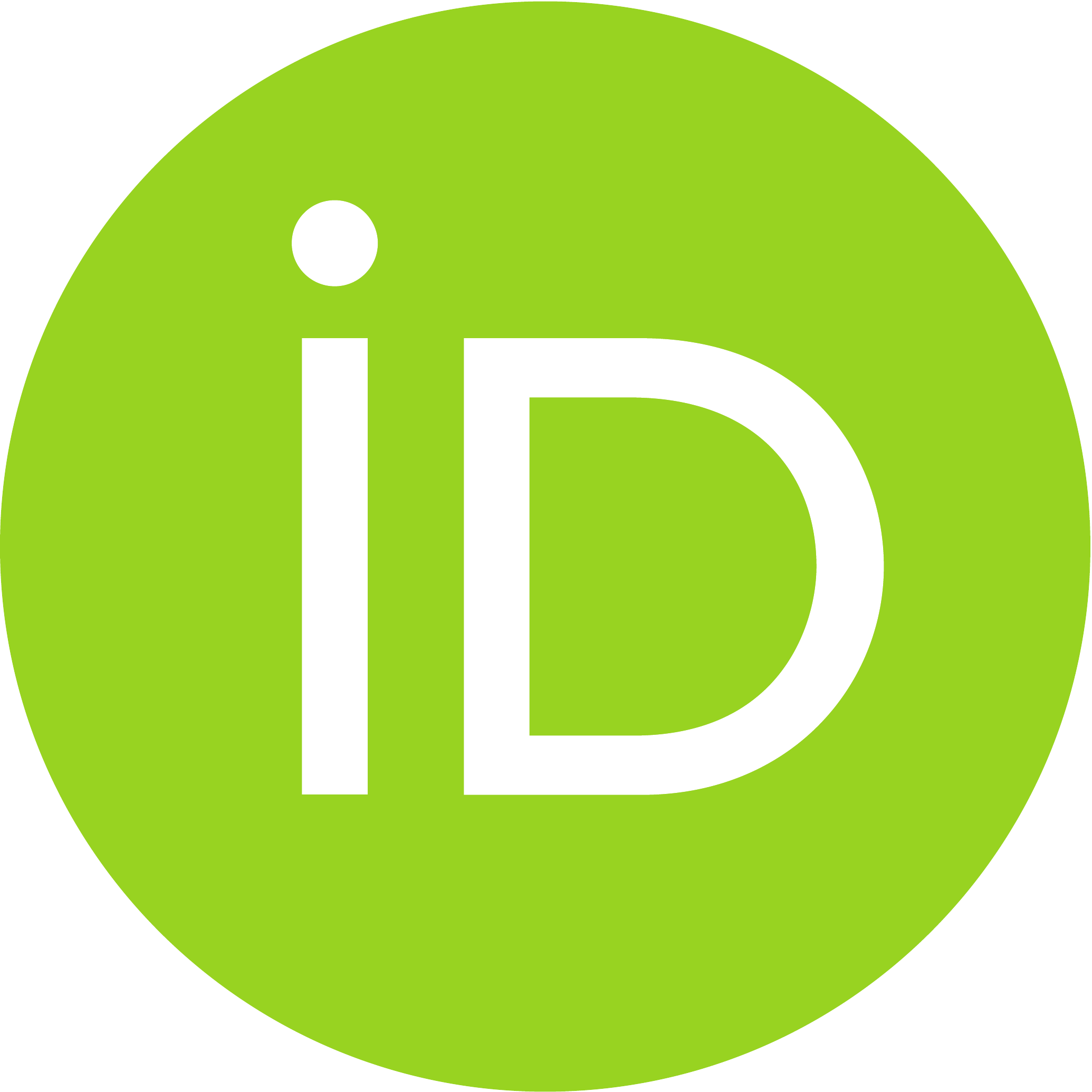}~https://orcid.org/0000-0001-8405-0141} and Pierre-Yves Longaretti\sup{1,2}\thanks{\includegraphics[width=0.3cm]{orcid.pdf}~https://orcid.org/0000-0002-4940-0756}}
	\affiliation{\sup{1}Université Grenoble Alpes, CNRS, Inria, Grenoble INP, LJK, 38000 Grenoble, France}
	\affiliation{\sup{2}Université Grenoble Alpes, CNRS-INSU, IPAG, CS 40700, 38052 Grenoble, France}
	\email{francois-remi.mazy@inria.fr, pierre-yves.longaretti@inria.fr}
}


\abstract{The use of spatially explicit land use and land cover (LULC) change models is widespread in environmental sciences and of interest in public decision-help. However, it appears that these models suffer from significant biases and shortcomings, the sources of which can be mathematical, conceptual or algorithmic. We formalize a modeling environment that distinguishes a calibration-estimation module and an allocation module. We propose an accurate calibration-estimation method based on kernel density estimation and detail an unbiased allocation algorithm. Moreover, a method of evaluation of LULC change models is presented and allows us to compare them on various fronts (accuracy, biases, computational efficiency). A case study based on a real land use map but with known (enforced) transition probabilities is used. It appears that the estimation error of the methods we propose is substantially improved over the best existing software. Moreover, these methods require the specification of very few parameters by the user, and are numerically efficient. This article presents an overview of our LULC change modeling framework; its various formal and algorithmic constituents will be detailed in forthcoming papers.}

\begin{document}
\onecolumn \maketitle \normalsize \setcounter{footnote}{0} \vfill

\section{\uppercase{Introduction}}
\label{sec:introduction}

Land use and land cover (LULC) change is a major driver of global change alongside the more visible issues of climate change and biodiversity loss. The study of LULC change is of major interest in analyzing and understanding a variety of socio-environmental phenomena but also for decision-help on mitigation and adaptation policies, and the literature on LULC change studies is by now quite substantial. Different LULC change modeling strategies have been developed over the last few decades to address these research issues (static, dynamic, agent-based, local or global scale, \emph{etc}\dots). 

In the present work, we focus on spatially explicit, statistical LULC change model building. Such models generally consists of estimating transition probabilities from one land-use state to another based on (usally) two  land-use maps at different dates that reflect changes in the past. These models are designed to simulate new land use maps, in scenario-driven projections of future land use and cover. Such allocated maps can be used for a variety of purposes. For example, they may be coupled to models of ecosystem services to produce evaluations of their future evolution. In fact, producing accurate spatially explicit projections of the effects of public decisions bearing on social-ecological problems is a current issue for such models \cite{Verburg2019a}.

Statistical and spatially explicit LULC change models constitute a popular approach to LULC change modeling, and various model-building software have already been devised in this framework. Among the most well-known, one may quote Dinamica~EGO \cite{SoaresFilho2002}, Idrisi LCM \cite{Eastman1995}, or the CLUE family of models \cite{Verburg1999}, the last member being CLUMondo \cite{Vliet2015}.
However, these software (and others) exhibit substantial differences in results for the same case studies \cite{Mas2014, Prestele2016, Alexander2017}, which raises an important concern on the reliability of the LULC change modeling process itself. 

No study to date has clearly identified the origin of the differences of behavior and outcome displayed by the existing model-building software on a same problem, let alone proposed a formally correct theoretical framework for this type of LULC change modeling. Moreover, comparing different LULC change models on their spatial outcomes is a poorly mastered operation in the literature \cite{Vliet2016} and more precise evaluation methods are needed. Nevertheless, as will be illustrated in the course of this work for the three modeling environment just mentioned, these discrepancies can be traced to differences in the formal and algorithmic choices made in their elaboration. In fact, quite a few of these choices were made without paying sufficient attention to the constraints imposed by a correct \textit{ab initio} formal investigation of the probabilistic foundation of the problem, leading to a number of errors and biases. 

More precisely, we address these issue by focusing successively on the calibration-estimation and allocation processes in order to propose a fully formalized and conceptually correct foundation for statistical spatially explicit LULC change analyses. We also provide efficient and bias-free algorithmic implementations of these processes, which constitute the core of this approach to LULC change model-building. The present work illustrates some of our main results on these two fronts. A detailed exposition of the underlying formal analyses and algorithmic implementations will be given in forthcoming dedicated papers. The first one will formalize in a rigorous ways statistical LULC changes by groups of contiguous pixels (patches), starting from the more common pixel by pixel probability distributions, and introduce an allocation algorithm to this effect, that will be explicitly shown to be bias-free. The second one will be dedicated to the calibration and estimation of probability transitions. Finally, the last article will provide a systematic method of identification and a systematic analysis of the biases present in the software mentioned above.

\begin{figure}
	\centering
	\includegraphics[width=0.92\linewidth]{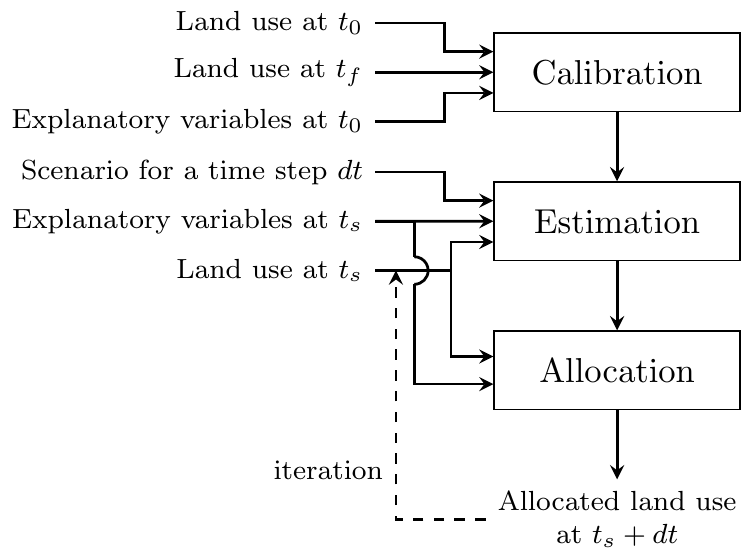}
	\caption{General architecture of a spatially explicit statistical LULC change model.}
	\label{fig:architecture_lucc}
\end{figure}

This paper is organized as follows. The typical structure of statistical spatially explicit LULC change models is recalled in section~\ref{sec:problem_formulation}. Next, a new and accurate calibration-estimation method and an unbiased allocation methods are briefly overviewed in section~\ref{sec:methods}. We then propose an evaluation method for LULC change models that allows us to compare their results and performances; the most salient features are described in section~\ref{sec:evaluation_methods}. The next section introduces a case study to illustrate the methods of the previous sections, and presents some of our results (section~\ref{sec:results}). Finally, the last section summarizes these findings along with other important results on these issues that will be established in our forthcoming papers on the topic.

We assume that a number of preliminary modeling choices are made prior to the comparison of performances mentioned right above: elaboration of the LUC typology, choice of explanatory variables, and choice of the discretization (pixel) scale. Thus, we exclude from the perimeter of our analysis the obvious influence of these choices on the results \cite{GarciaAlvarez2019}. 

\section{\uppercase{LULC change model architecture}}
\label{sec:problem_formulation}

Spatially explicit statistical LULC change models are generally (but not always) organized around two main modules: calibration-estimation, and allocation  (Fig.~\ref{fig:architecture_lucc}). This allows us to investigate in turn their underlying logic, formal foundation and algorithmic implementation.

\subsection{Calibration-Estimation Module}
\label{sec:def_calibration}
We start from two raster (i.e., pixelized) LULC maps at two different dates $t_0$ and $t_1$ as well as maps of the $d$ explanatory variables of interest, characterizing LULC changes (e.g., altitude, distance to the roads, \emph{etc}\dots). Pixel sizes can vary from a few meters to a few tens of kilometers depending on case studies; the largest maps may involve tens of millions of pixels. The dates $t_0$ and $t_1$ differ typically by a few years. 

Although changes occur in patches of contiguous pixels, the calibration-estimation process focuses on individual pixel probability distributions; patches are then produced on this basis in the allocation module. Pixels undergo a change from land use $u$ to land use $v$. Each pixel $i$ is characterized by a collection (vector or tuple) $\mathbf{z}_i$ of values of the $d$ explanatory variables. The transition probability distribution must be evaluated for all possible explanatory variables combinations $\mathbf{y}$ present in the calibration maps, including those which were not observed to undergo a transition but may be expected to in the future. To this effect, one needs to evaluate the conditional pixel transition probability $P(v|u,\mathbf{y})$. This involves a form of interpolation in explanatory variable space, called estimation (Fig.~\ref{fig:architecture_lucc}).  

\subsection{Allocation Module}
\label{sec:def_allocation}

A new land use map is simulated from an initial one (possibly different from the calibration maps), from the explanatory variables (when needed) and from the probability of land use change maps for each possible LULC state change  (Fig.~\ref{fig:architecture_lucc}). Allocations are made in groups of contiguous pixels (patches). The process involves a sampling algorithm and attributes a possible new LULC state on the basis of the associated transition probability distribution and some patches constraints. The map obtained in this way is a particular instantiation of this probability distribution; many others could have been produced for the same time step, and all such maps are statistically equivalent, except for statistical noise. This point is usually ignored in existing model-building environments, although it is crucial to identify errors and biases, as will be discussed later on. The details of our patch construction method are involved but not critical for software performance comparison (both in terms of accuracy and execution time) and will therefore not be presented here.

In the process the user provides a transition matrix that characterizes the rates of change per time step between each land use state, and characterized by an overall probability $P(v|u)$ for all possible LULC initial and final states $u,v$. This matrix is usually specified through scenarios. 

\subsection{Existing LULC change Model-Building Environments}
\label{sec:existing_models}

In the present work we focus on three of the most widely used software~\cite{Vliet2016}: Dinamica~EGO, Idrisi~LCM and CLUMondo. Our objective is to contrast their performances (accuracy and computational efficiency) to the modeling strategy and algorithmic implementation we propose in the next section.

Dinamica~EGO fits reasonably well in the archetypal structure of Fig.~\ref{fig:architecture_lucc}. Transition probabilities are usually estimated from the use of weights of evidence, applied after binning the user-defined explanatory variables \cite{SoaresFilho2002}. This binning step requires to specify a number of parameters, which turn out to have a very strong impact on the obtained results. Moreover, this method of estimating transition probabilities is based on the strong assumption of statistical independence of the explanatory variables; it turns out that even modest amounts of cross-correlation between these variables may lead to non negligible errors in the estimated transition probabilities. This software implements an allocation method relying on a pixel pre-selection process (pruning), implemented to reduce computational time); such pruning also produces significant biases in the results.

Idrisi LCM \cite{Eastman1995} also implements a calibration/allocation architecture, while leaving the user with less control on modeling choices than Dinamica~EGO. This strategy is adopted on purpose, in order for users with little expertise to nevertheless be able to implement a LULC change model. The particular LCM version we have used is the one bundled in Idrisi Selva. This software proposes different models for estimating transition probabilities: logistic regression (LCM LR), SimWeight (LCM SW) and Multi-Layer Perceptron (LCM MLP). The LCM allocation module is very simple and deterministic, as the allocation algorithm implemented in this software essentially ignores the statistical nature of the process. This results in simple but strongly biased allocation rules.

Finally, the CLUE family of software \cite{Verburg1999}, the latest of which is CLUMondo \cite{Vliet2015}, chooses to estimate the probabilities of change from a single land use map, from which it defines transition probabilities through a logistic regression on all explanatory variables. Also, CLUMondo does not allow the modeler to access its allocation module independently.

\section{\uppercase{CLUMPY: A New LULC change Model-Building Environment}}
\label{sec:methods}

We present here in a succinct manner the methods used in our calibration-estimation and allocation modules. The closest analog in the existing model-building environments is Dinamica EGO. The main differences come from our more sophisticated but more accurate and efficient calibration-estimation method, our demonstrably error- and bias-free allocation method, and significantly more efficient algorithms leading to substantial gains in computation time in large problems.

These modules constitute the core of our model-building environment, called CLUMPY (Comprehensive Land Use Model in Python).

\subsection{Calibration-Estimation}
\label{sec:methods_calibration}
\subsubsection{Bayes Rule}

The probability of transition from land use $u$ to land use $v$ of a pixel characterized by the $d$-tuple of explanatory variables $\mathbf{y}$ is noted $P(v|u,\mathbf{y})$. This notation emphasizes the fact that this probability is conditioned by the knowledge of the values of $u$ and $\mathbf{y}$. Bayes rule allows us to express this probability in a more convenient way for calibration purposes:
\begin{align}
P(v|u,\mathbf{y}) = P(v|u) \cfrac{p(\mathbf{y}|u,v)}{p(\mathbf{y}|u)}.
\label{eq:bayes}
\end{align}

\noindent The various factors on the right-hand side member are defined as follows. $P(v|u)$ is the global transition probability, specified by a scenario provided by the user. This quantity can also be computed directly from the observed transitions between the calibration maps at times $t_0$ and $t_1$. Next, $p(\mathbf{y}|u)$ is a conditional probability density. We use probability densities because explanatory variables are usually continuous quantities, and we do not bin them in the end, in order to produce a more accurate calibration procedure. By definition $p(\mathbf{y}|u)d\mathbf{y}$ is the probability of observing $\mathbf{y}$ within the small volume $d\mathbf{y}$ of explanatory variable space, for pixels of initial state $u$. Similarly, $p(\mathbf{y}|u,v)$ is the probability density of $\mathbf{y}$ for pixels making the transition from states $u$ to $v$. Note that the calibration process, which is based on observed transitions, uses these probabilities in the frequency of occurrence meaning, while the allocation module applies Bayes rule in a Bayesian sense.

The interest of using Bayes rule lies in the fact that $p(\mathbf{y}|u)$ and $p(\mathbf{y}|u,v)$ are much simpler to estimate than $P(v|u,\mathbf{y})$. Indeed, it is then a question of estimating density functions on $\mathbf{y}$ for a set of pixels of initial state $u$ for $p(\mathbf{y}|u)$ or potentially undergoing a LULC state transition from $u$ to $v$ for $p(\mathbf{y}|u,v)$. This problem is widely addressed in machine learning and is called density estimation.

Consider a set of $n$ calibration pixels, made of all the pixels that actually underwent a LULC state change from $u$ to $v$ and are directly extracted from two calibration maps at time $t_0$ and $t_1$. Each of these pixels is characterized by its explanatory variables noted $\mathbf{z}$. We wish to estimate the transition probability of all $m$ pixels in our maps; these are by definition evaluation pixels. Each of these pixels is characterized by its explanatory variables noted this time $\mathbf{y}$. The problem is therefore to estimate $P(v|u,\mathbf{y})$.

The idea is to calibrate a density estimator with the calibration pixels for both probability densities and then to estimate these probability densities for all pixels with the obtained density estimator. Finally, as the global transition probability $P(v|u)$ is given, we can apply Bayes rule to obtain $P(v|u,\mathbf{y})$.

\subsubsection{Density Estimation by Kernel Density Estimator (KDE)}\label{sec:kde}

Estimating a probability density is a widely addressed problem in the machine learning literature \cite{Scott2015}. A simple and not very precise method is to make histograms in explanatory variable space. This requires to bin explanatory variables; unfortunately, the choice of the bin size has a strong impact on the results. Some related methods such as averaged shifted histograms have been proposed to circumvent the problem \cite{Chamayou1980}. 

More sophisticated methods perform density estimates by positioning a kernel function $K$ characterized by a user-specified width parameter $h$ (called bandwidth) on each calibration data point and summing over all these kernels at the desired locations of estimation (in explanatory variable space). This is called kernel density estimation \cite{Wand1992}, or KDE in short. This very efficient method is however highly computationally intensive as soon as the number of explanatory variables and elements (here, pixels) increases. Many kernel functions may be studied (e.g., gaussian kernels), and various methods have been proposed to approximate the resulting family of estimators \cite{OBrien2016, Charikar2017}. However, these relatively complex methods often require large amounts of memory since they allocate full matrices representing the entire space of explanatory variables. Thus, when confronted with more than a few explanatory variables, these methods can be inapplicable due to lack of memory space on standard machines; the computation time also increases quite fast with the number of dimensions.

Instead, we implemented a hybrid binning/KDE method, that keeps part of the simplicity of simple histograms with only a small degradation of the performance of pure Kernel density estimator methods. The binning is performed on a scale $h/q$ smaller than the bandwidth $h$ \cite{Wells2019}, where $q > 1$ is an odd integer. A kernel density estimator is then applied to the small bins themselves instead of the original individual elements (pixels) in explanatory variable space. One can then show that the larger $q$, the smaller the approximation error. This hybrid method maintains the spirit of continuous density estimation, while being much more computationally efficient than direct KDE estimators (both in terms of computational time and memory use). This is what we used in our estimations of $P(\mathbf{y}|u)$ and $P(\mathbf{y}|u,v)$. 

The choice of the kernel width $h$ for the KDE method is a very important issue due to its influence on the quality of the estimation. Indeed, choosing a too narrow bandwidth results in an over-interpretation of the observational data and in noisy estimates. On the other hand, a too broad bandwidth leads to under-fitting (or over-smoothing) and to a degradation of the estimate of the probability density. The determination of the optimal width of the kernel is a non-trivial problem and is widely discussed in the machine-learning literature \cite{Wand1994,Rudemo1982,Sain1994}. However, these methods are computationally very expensive as soon as the number of pixels is large and the number of dimensions exceeds 3, which is very frequently the case in LULC change studies.

This being said, in the LULC change context, a slight over-smoothing is not a problem and can even be interesting, because transition probabilities are usually undersampled, and therefore noisy. Consequently, we chose to determine the KDE bandwidth $h$ from the principle of maximum smoothing of Terrel \cite{Terrell1990}, leading to
\begin{align}
h_\text{Terrel} = \cfrac{(d + 8)^{\frac{d+6}{2}}~ \pi^{\frac{d}{2}} \displaystyle\int K^2}{16~n~(d+2)~\Gamma\left(\frac{d+8}{2}\right)},
\end{align}
where $K$ is the kernel density estimator. The slight oversmoothing involved turns out to be essentially unnoticeable in our tests, while this prescription considerably reduces the computational burden of KDE methods. 

\noindent We have checked on various test problems, and will soon show, that even simple kernel functions (such as a square box, a triangle or a gaussian kernel) give much more accurate results than existing calibration methods.

Note finally that, because explanatory variables are of widely different nature and not statistically independent, it is necessary to normalize them in order to work with data of zero mean and covariance matrix equal to the identity in explanatory variable space. This operation is called ``whitening transformation'' in the machine learning literature. It makes it legitimate to use a unique bandwidth in all dimensions in the transformed explanatory variable space and greatly simplifies the numerical implementation of our calibration-estimation method.

\subsection{Allocation}
\label{sec:methods_allocation}

In general, the allocation module takes as input a land use map as well as the transition probability maps obtained in the calibration-estimation module. The allocation method presented here also requires the knowledge of the explanatory variables of the input map.

The simulation of an allocated map produces a specific statistical sampling of the transition probabilities. So far, we have focused on pixel transition properties. However, as already mentioned, we allocate pixel patches. Following the logic initiated in Dinamica EGO, a first pixel is selected according to the transition probability distribution obtained by our calibration-estimation procedure. This pixel is called a ``core pixel'' or ``pivot-cell''. Then, a specific procedure is applied to create a patch around this core pixel. This procedure is characterized by different parameters such as the surface of the patch, and its elongation. 
Patches created in this way reproduce some of the statistical properties of actually observed transition patches, but are defined algorithmically to provide some randomness in their shape.

The allocation modules implemented in the existing software turn out to be all biased to various extents. Such biases have not yet been identified in the literature, first because a biasing criterion has not been formulated, and second because the details of these software allocation procedure is not fully documented. We have circumvented this last problem by a combination of literature analysis, educated guesses, retro-engineering, questions to model developers, and re-implementation (when feasible) of these allocation algorithms to check that our understanding of their structure and content exactly reproduces the outcome of the original software on a number of test problems. We have also formulated a simple but powerful ``no-bias'' criterion, which, in essence, requires that the post-allocation probability distributions are identical to the pre-allocation ones. This is in fact more of an unavoidable self-consistency requirement, but it turns out that none of the existing LULC change model-building software does satisfy it.

This allowed us to implement a strategy of systematic identification and characterization of the various biases present in existing model-building software. We illustrate this process on a particularly important bias related to pruning, relying on an efficient bias-free algorithm which avoids the need for pruning.

\subsubsection{Pruning}
\label{sec:pruning}

LULC change models may involve a very large number of pixels (e.g., tens of millions). Therefore, it can be interesting to pre-select a limited number of pixels in order to speed up the allocation procedure. This pre-selection is called pruning and is implemented by Dinamica EGO and LCM in two different ways, both of which turn out to be significant sources of bias.

Dinamica EGO's pruning method consists in ranking pixels by decreasing order of transition probability $P(v|u,\mathbf{y})$. Pixels are then pre-selected in this order in this list; the number of pre-selected pixels is equal to the number of pixels necessary to reach the targeted LULC change surface defined by the user selected scenario, multiplied by a pruning parameter $F$. The default value is $F=10$ (ten times as many pixels as needed for the various transitions are pre-selected). 

LCM ranks the pixels by decreasing order of the probability density (or probability distribution if binned) of the explanatory variables for this transition, $p(\mathbf{y}|u,v)$. Then LCM keeps the exact number of pixels that are necessary to reach the targeted transition surface defined by the selected scenario, and has a specific procedure to resolve conflicts of allocation for the same pixel. Thus, LCM's pruning method is also its allocation method since all pixels selected in this way are directly allocated without any further consideration. This procedure has the advantage of simplicity and ensures that transitions occur in patches (due to the spatial continuity of the explanatory variables probability densities). However, the maps allocated by LCM often suffer from a severe lack of realism, and they always violate our self-consistency no-bias requirement.

Although establishing which probability ordering should in principle be used for pruning is not an obvious task, we have proved that LCM's choice [$p(\mathbf{y}|u,v)$] is the theoretically correct one; this applies although the correct allocation probability distribution is by definition $P(v|u,\mathbf{y})$. This being said, both Dinamica~EGO and LCM pruning strategies are strongly biased because they modify to various extents the probability distributions which should be enforced exactly. LCM is the more biased of the two, although Dinamica EGO has chosen an incorrect probability ordering for pruning.

An unbiased pruning method necessarily consists in a random sampling performed according to $p(\mathbf{y}|u,v)$, and selects the number of kernel pixels needed to reach the targeted number of transited pixels. Thus, the probability distribution of this pixel subsample will be statistically representative of $p(\mathbf{y}|u,v)$ and there will be no unwarranted truncation of this probability density distribution. 

The motivation for applying a pruning procedure lies in the numerical acceleration of the allocation method for a very large number of pixels. However, the implementation in Python of the our allocation procedure as presented in section~\ref{sec:unbiased_allocation} proves to be numerically very efficient without pruning, even though the number of pixels in some of our case studies is very large ($>10^8$); this efficiency relies on the use of dedicated Python functions, which perform nearly as fast as equivalent C codes. Still, we have designed a bias-free pruning algorithm, for possible use in particular problems.

\subsubsection{Unbiased Allocation}
\label{sec:unbiased_allocation}

In addition to pruning, other biases can occur in the allocation procedure. In particular, the creation of patches around a core pixel automatically excludes these patch pixels from the rest of the allocation procedure. However, they could just as well be selected afterwards. 

In order to resolve such potential conflicts, we designed the following  iterative allocation procedure for a given initial land use state :

\begin{enumerate}
	\item\label{alloc:gart} We know the transition probability $P(v|u,\mathbf{y})$ of each pixel for all ($u,v$). We can therefore apply a generalized Von Neumann rejection sampling, which allows us to test all possible states $v$ at the same time for any given $u$. We obtain an unbiased sample of kernel pixels for each of the transitions studied. If no kernel pixel is selected, the allocation procedure is terminated at this point.
	\item A single kernel pixel is randomly and uniformly drawn; its associated transition has been determined in the previous step.
	\item The procedure of patch creation around this core pixel is applied next. The selected pixels are actually allocated on the simulated map.
	\item\label{alloc:update_P_v__u} $P(v|u)$ is updated, taking into account that a certain number of pixels have already been allocated. This probability is therefore reduced for the rest of the allocation procedure in the considered time step.
	\item\label{alloc:update_P_y__u} $p(\mathbf{y}|u)$ is updated next because some elements have already been allocated, which has modified the probability density distribution of the explanatory variables.
	\item $p(v|u,\mathbf{y})$ is then recalculated from Bayes rule Eq.~\eqref{eq:bayes}; the probabilities involved in steps \ref{alloc:update_P_v__u} and \ref{alloc:update_P_y__u} are also updated. We then start again at step \ref{alloc:gart}.
\end{enumerate}

We have shown that a carefully designed algorithm of this type is bias-free.
This procedure is however demanding since it requires to re-estimate very frequently $p(\mathbf{y}|u)$. We therefore update this probability distribution only when the percentage of allocated pixels is significant enough that the estimated probability is too far from the real distribution (in practice, after a fixed small percentage of state changes has been achieved). This may introduce a slight bias.

\section{\uppercase{Evaluation}}
\label{sec:evaluation_methods}

A review of the literature shows that validating the results obtained by LULC change models is an uncommon practice. This deficiency underlies some of the doubts that may be raised on the robustness of LULC change modeling \cite{Vliet2016}. In any case, before specific results may be validated, the modeling framework itself should be validated. A first attempt along these lines has already been carried out on artifical data \cite{Mas2014} but the lack of exact knowledge of the transition probabilities involved precluded any detailed evaluation of the LULC change models that were tested in this earlier work. Indeed, one of the main problems of statistical LULC change modeling is the estimation of the probability distribution $P(v|u,\mathbf{y})$ (section~\ref{sec:methods_calibration}), and, furthermore, it is impossible to know this probability distribution exactly in a real case study. Validating this estimation of $P(v|u,\mathbf{y})$ therefore remains an essential objective.

We propose here a simultaneous validation method for the calibration-estimation and global calibration-estimation/allocation procedures, that allows us to objectively compare different model-building strategies. This is achieved by quantifying the difference between the transition probabilities obtained from our hybrid KDE estimation in the calibration phase or in the  post-allocation one, and exactly specified pre-calibration transition probabilities. We may proceed to this effect from semi-real or completely artificial data. We start from a (real or artificial) LULC map at $t=t_0$. We adopt an exactly known transition probability distribution, $P^*(v|u,\mathbf{y})$. This exact transition probability may be defined analytically or numerically. Then, a new LULC map at time $t=t_1$ is produced with our allocation procedure (based on this exact probability distribution and our patch creation algorithm). 

This allows us to implement two validation processes: 

\begin{description}
	\item[Calibration-Estimation comparison] We select a calibration-estimation method from an existing LULC change model-building environment, and produce from the two maps just specified at $t_0$ and $t_1$ an estimate $\hat{P}(v|u,\mathbf{y})$ of $P^*(v|u,\mathbf{y})$. We repeat this process for all the calibration methods we want to compare, including ours.
	\item[Calibration-Estimation/Allocation comparison] We now wish to  evaluate the relative efficiencies of these modeling environments over the whole calibration-estimation/allocation process. We thus use as inputs the LULC map at $t=t_1$ and the probability distribution $\hat{P}(v|u,\mathbf{y})$ determined by the previously described  calibration-estimation comparison process, for any of the modeling environments tested. We produce next a new LULC map at $t=t_2$ from the associated allocation procedure. We recover a new post-allocation estimate $\tilde{P}(v|u,\mathbf{y})$ of $P^*(v|u,\mathbf{y})$ from the calibration-estimation method of the same modeling environment. We repeat this process for all the modeling environments we want to compare, including ours.
\end{description}

The comparison of the exact (enforced) probability distribution with the estimated ones can be done in various and more or less sophisticated ways. In this article, we limit ourselves to two very simple approaches. The first one consists in producing graphs of the estimated distributions $\hat{P}(v|u,\mathbf{y})$ and the exact one $P^*(v|u,\mathbf{y})$ considered as a function of $\mathbf{y}$, for various one-dimensional cuts in explanatory variable space, e.g., by fixing all explanatory variables but one. This gives a direct check of the accuracy of the various methods, but only on a limited (although randomly chosen) set of locations in explanatory variable space.

The second validation method is more global and consists in calculating the average of the absolute error throughout explanatory variable space:
\begin{align}
\varepsilon_{\text{calib}} & = \frac{1}{m} \sum_{i=1}^m \left| P^*(v|u,\mathbf{y}^i) - \hat{P}(v|u,\mathbf{y}^i) \right|,\nonumber\\
\varepsilon_{\text{tot}} & = \frac{1}{m} \sum_{i=1}^m \left| P^*(v|u,\mathbf{y}^i) - \tilde{P}(v|u,\mathbf{y}^i) \right|
\label{eq:mae}
\end{align}
where $m$ is the number of pixels where the difference is evaluated (generally, all pixels concerned by the $u \rightarrow v$ transition in a map), and where the subscript ``calib'' or ``tot'' refers either to the calibration-estimation comparison process or to the global calibration-estimation/allocation one.

By construction Eq.~\eqref{eq:mae} tends to underestimate large but localized differences. Using one-dimensional cuts minimizes this possibility to some extent. We could also, e.g., identify the largest absolute difference, and count the number of pixels where this difference is achieved within a given tolerance. We avoid relative differences because they might be very large where the transition probability is low, but this would not necessarily reflect a notable inaccuracy in the estimation itself.

In any case, we can measure the difference between the exact and estimated probability distributions, a validation test that has been never been performed so far, and thus have a first global evaluation of the relative robustness of various calibration-estimation and allocation modules (ours as well as the ones implemented in existing model-building software).

Our own calibration-estimation and allocation procedures are tested simultaneously and not independently in both comparison processes (as the second map needed in these tests is produced by our allocation procedure). One may therefore ask whether they are both validated in this way. Several lines of arguments show that this is the case, relying on \textit{a priori} and \textit{a posteriori} analyses. First, we have formally proved  that our allocation procedure is bias-free (the proof will be given elsewhere), i.e., that it enforces the correct transition probability distribution. Also, the KDE density estimation procedures have been shown to converge exactly to the correct density distribution in the machine-learning literature in the limit of a large number of available points (this form of convergence is weaker than for allocation, but still relevant). This applies also to our own hybrid KDE calibration-estimation method, except possibly close to boundaries in explanatory variable space due to the kernel truncation correction applied there (not described in our overview of the method). Second, the numerical efficiency of our algorithms allow us to produce a large  number (thousands) of calibration-allocation sequences for the same time step in a reasonable computation time. We checked on a number of test problems that post-allocation transition probabilities (recovered by our calibration-estimation procedure) converge in expectation value to the enforced one by averaging over these multiple allocations. These arguments provide an \textit{a posteriori} validation of both procedures. Indeed, having errors or biases introduced by one procedure nearly exactly compensated by the other is beyond unlikely, considering the very different strategies used in their elaboration.

A similar concern may be raised about the fact that using our own allocation procedure to produce the $t=t_1$ LUC map for testing calibration-estimation procedures may favor our method over the others. This concern is misplaced for the same type of reason: the way this second map was produced is irrelevant precisely because there is no relation between our allocation procedure and any of the calibration-estimation procedures we have tested. Also, we have just shown (or at least convincingly argued, until the above-mentioned proofs are available in print), that our allocation procedure produces unbiased post-allocation probability distributions, so that the amount of bias (due in fact to statistical noise) produced on a single $t=t_1$ map is limited. In any case, it applies equally to all tested calibration methods, which are therefore treated on the same footing in this respect.

\section{\uppercase{Results and Discussion}}
\label{sec:results}

In this section we apply the evaluation methods presented in section~\ref{sec:evaluation_methods}. We thus define a case study that is intended to be representative of commonly encountered LULC change problems. We put to the test our own modules (section~\ref{sec:methods}) as well as the ones implemented in the existing model-building software introduced in section~\ref{sec:existing_models}. The parameters used for each model are specified in the Appendix. 

\subsection{Case Study Short Description}
\label{sec:case_study_def}

We are interested in a study area of 94 square kilometers located in the Isère \textit{département} in France. We focus on a smaller sector in the Southwest of the town of Grenoble; we have raster maps describing this smaller area at 15 meters of resolution (6.3 million pixels), and use 7 different land use classes at the coarsest typology level (water bodies, mineral areas, forests, agricultural areas, urban areas, economic activity areas and other), and up to several tens at the finest level. These data have been used in a recent project, whose objective was to explore the future of ecosystem services at the 2040 horizon under various land planning scenarios \cite{VLLL16,VBLNCCBQLL19,VLCBLCLL19}.

For our present purposes, we only focus on a single transition, namely  from agricultural areas ($u$) to urban areas ($v$). This is one of the main transitions responsible for urban sprawl. We chose this transition for the sake of clarity and simplicity. The number of agricultural area pixels is $\sim 3.3$ million.

We have selected three explanatory variables to characterize this transition: elevation above sea level in meters ($y_0$), slope in degrees ($y_1$) and shortest distance from urban areas in meters ($y_2$). These are three of the main explanatory variables typically used in urban sprawl studies relying on statistical LULC modelling frameworks.

In line with the evaluation strategy described in section \ref{sec:evaluation_methods}, we enforce a specific transition probability distribution, namely we adopt a multivariate Gaussian distribution: 
\begin{align}
P^*(v|u,\mathbf{y}) =
\left\{
\begin{array}{ll}
\mathcal{N}_{\mu,\Sigma}(y_0, y_1, y_2) & \mbox{ \text{if}~$\{y_0, y_1, y_2\}\in \mathcal{D}$} \\
0 & \mbox{ \text{else}} 
\end{array}
\right. 
\label{eq:f}
\end{align}
\noindent where $\mathcal{N}_{\mu,\Sigma}$ refers to a normal distribution of mean $\mu$ (vector of the means of the explanatory variables), and covariance matrix $\Sigma$ (covariance matrix of the explanatory variables), $(y_0, y_1, y_2)$ is the vector of explanatory variables, and $\mathcal{D}$ is a subset of the explanatory variable space where the probability density $P(\mathbf{y}|u)$ is larger than a (small) threshold (this avoids potential problems in the application of Bayes rule without introducing any significant bias in the transition probability distribution). The exact definition of $\mu$, $\Sigma$ and $\mathcal{D}$ are given in the Appendix.

\subsection{Mean Absolute Error Comparison}
\label{sec:results_mae}

The pixel-averaged absolute errors are calculated with Eq.~\eqref{eq:mae} of section~\ref{sec:evaluation_methods} and reported in Table~\ref{tab:mae}.

CLUMPY's $\varepsilon_\text{calib}$ is about four times lower than for the next best existing software with respect to this evaluation criterion, Dinamica EGO. This confirms the quality of the KDE estimator relative to other methods. As expected, $\varepsilon_\text{tot}>\varepsilon_\text{calib}$ whatever the model since the allocated map is a particular instantiation of the estimated probability distribution, which is itself an approximation of the exact one. We notice also the influence of the pruning factor $F$ of Dinamica EGO: reducing this parameter results in a larger error (note that $F=10$ is the default value of this parameter). This finding is consistent with the fact that Dinamica EGO's pruning procedure performs a sharp truncation of the probability density (section~\ref{sec:pruning}).

We can also repeat the allocation step in the combined calibration-estimation/allocation process. We have chosen to run it 100 times (last column of Table~\ref{tab:mae}), and average over these various runs in order to improve the precision of the estimation of the transition probability distribution. This is ineffective for LCM ($\varepsilon_\text{tot, 100}=\varepsilon_\text{tot}$), consistently with the fact that LCM allocation procedure is deterministic (see section~\ref{sec:pruning}). For Dinamica EGO, the improvement is marginal, which reflects the error due to pruning.
On the other hand, CLUMPY displays a significant improvement and comes very close to the value obtained for the calibration-estimation comparison: $\varepsilon_\text{calib} \approx \varepsilon_\text{tot,100}$. The allocation error itself has become negligible compared to the calibration-estimation one, consistently with the fact that our allocation algorithm is bias-free.

\begin{table}
	\caption{Comparison of models through Eq.~\eqref{eq:mae} for the  calibration-estimation and calibration/allocation comparison processes. DE = Dinamica EGO, for two different pruning factors ($F=10$ and $100$). The last column results from an average over 100 allocations of the same time step.}
	\label{tab:mae}
	\centering
	\begin{tabular}{l|c|c|c}
		model & $\varepsilon_\text{calib}$ & $\varepsilon_{\text{tot}}$ & $\varepsilon_{\text{tot, 100}}$\\
		\hline
		\vspace{-0.3cm}& & & \\
		CLUMondo & $3.56e^{-3}$ & & \\
		LCM LR & $3.35e^{-3}$ & $6.59e^{-3}$ & $6.59e^{-3}$\\
		LCM SW & $1.75e^{-3}$ & $5.85e^{-3}$ & $5.85e^{-3}$\\
		LCM MLP & $1.38e^{-3}$ & $6.58e^{-3}$ & $6.58e^{-3}$\\
		DE F=10 & $1.29e^{-3}$ & $5.31e^{-3}$ & $5.27e^{-3}$\\
		DE F=100 & $1.29e^{-3}$ & $1.65e^{-3}$ & $1.17e^{-3}$\\
		CLUMPY &$3.37e^{-4}$ & $9.84e^{-4}$ & $3.39e^{-4}$\\
	\end{tabular}
\end{table}

\subsection{One-Dimensional Cut Comparison}

\begin{figure*}
	\centering
	\begin{subfigure}[t]{.45\textwidth}
		\includegraphics[width=1\linewidth]{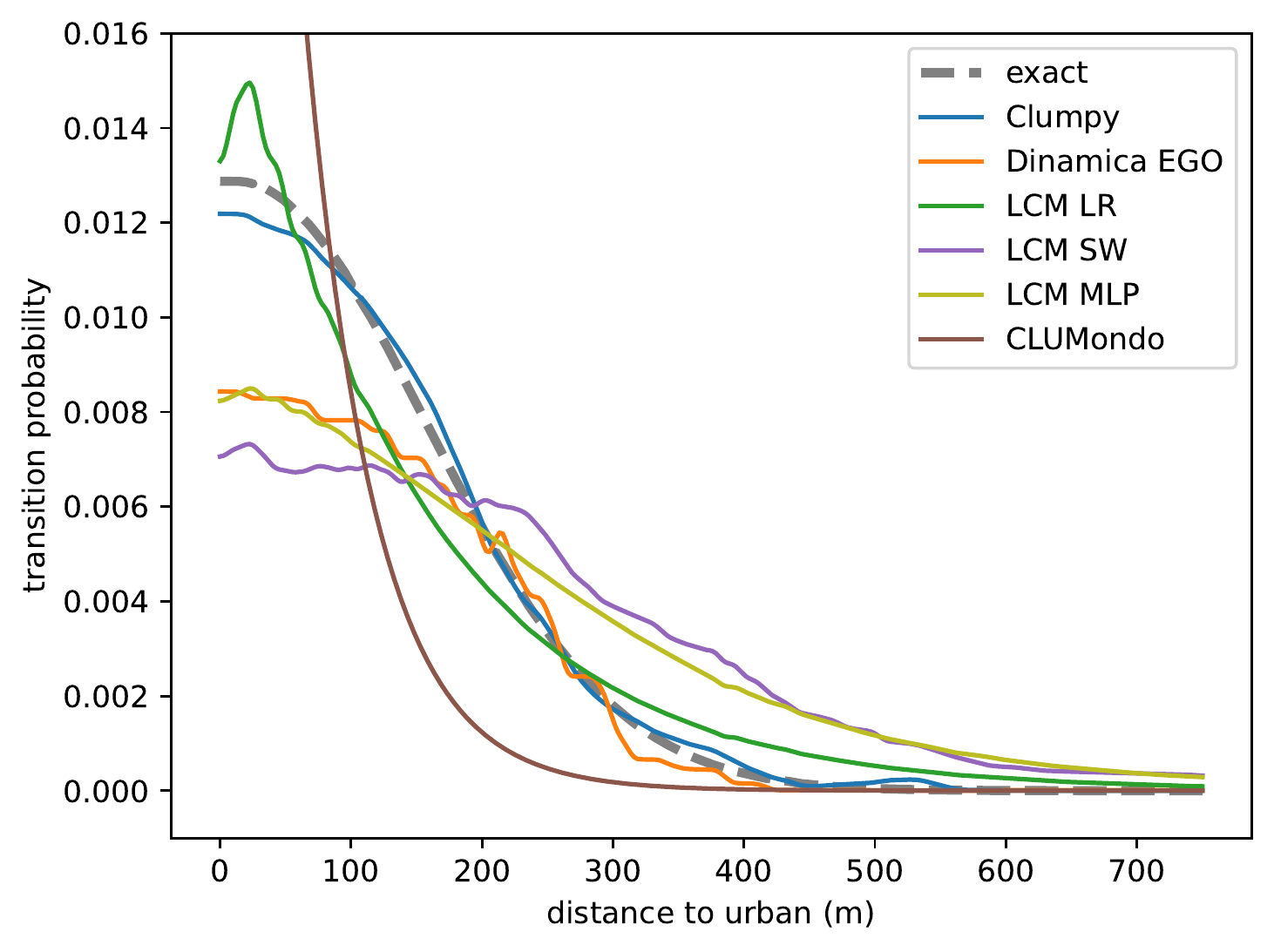}
		\caption{Calibration-estimation comparison process}
		\label{fig:cut_calib}
	\end{subfigure}
	\begin{subfigure}[t]{.45\textwidth}
		\includegraphics[width=1\linewidth]{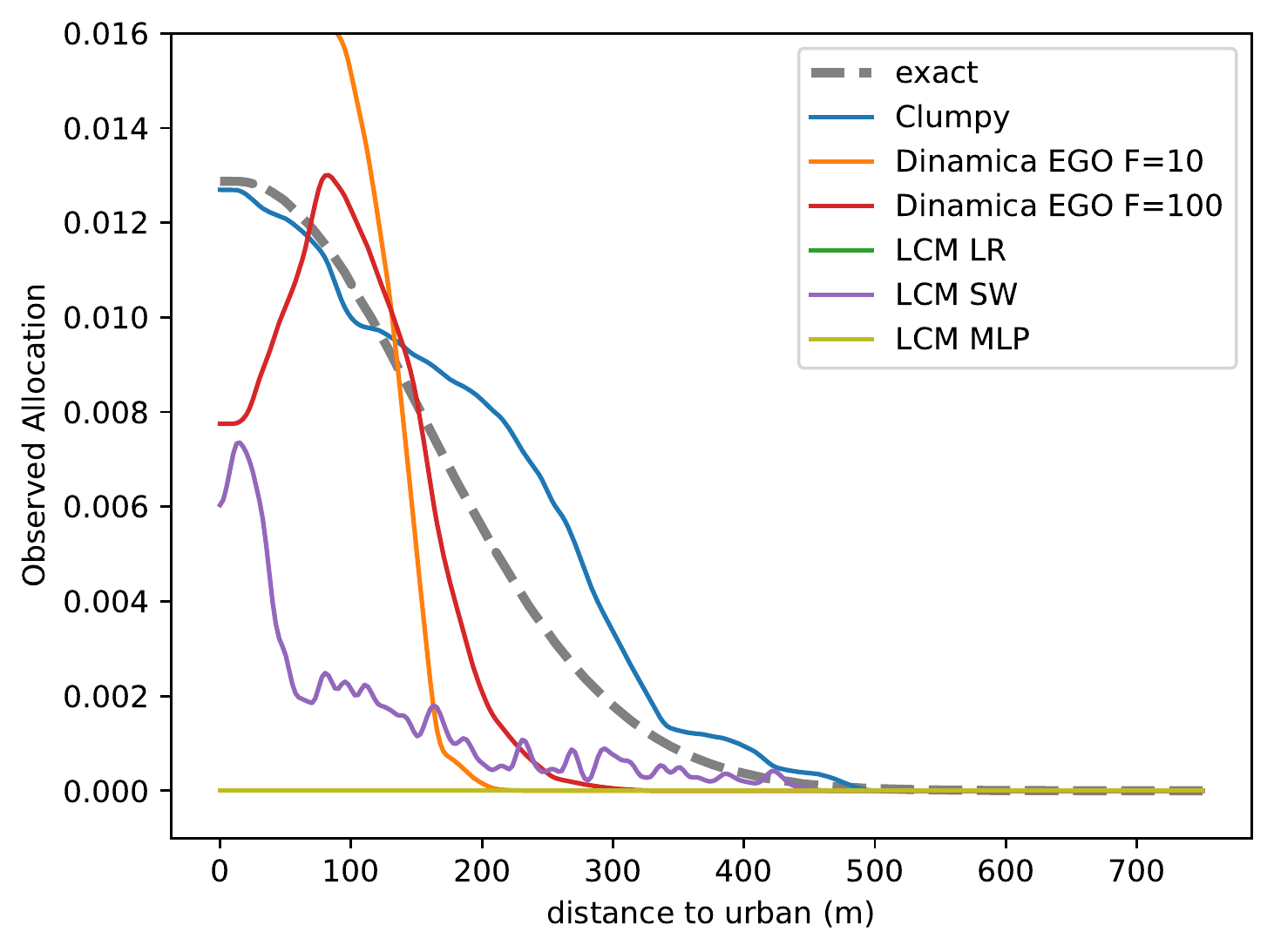}
		\caption{Global calibration/allocation comparison process (for a single allocation run)}
		\label{fig:cut_alloc}
	\end{subfigure}
	\caption{Transition probability one-dimensional cut from agricultural to urban areas with respect to distance to existing urban areas, with elevation set to $300~\text{m}$ and slope set to $2^o$.}
	\label{fig:cut}
\end{figure*}

The graphs of the estimated distributions $\hat{P}(v|u, \mathbf{y})$ and $\tilde{P}(v|u, \mathbf{y})$ returned by the different models are visible in Fig.~\ref{fig:cut}. This shows a one-dimensional cut at fixed altitude and slope, while using the shortest distance to existing urban areas as abscissa. The exact transition probability, computed from $P^*(v|u, \mathbf{y})$, Eq~\eqref{eq:f}, is also represented. There are two subplots, one for calibration-estimation comparisons [$\hat{P}(v|u, \mathbf{y})$], one for global calibration/allocation comparisons [$\tilde{P}(v|u, \mathbf{y})$]. Although this is a limited and local comparison, this exemplifies the behavior of each of the model-building algorithms.

Fig.~\ref{fig:cut_calib} represents the transition probabilities estimated by the calibration-estimation comparison process. We observe a large disparity in the obtained estimations. It seems fair to say that the existing modeling environments fail to represent in an accurate way the exact probability distribution, although it was chosen to be relatively smooth. The dispersion in these results must clearly contribute to the problem discussed in introduction, a point that will be quantified more precisely in the future. Our own modeling environment (CLUMPY) performs very well, comparatively and in absolute. 

Dinamica EGO displays a significant deviation from the exact curve which can be traced back to the assumption of independence of explanatory variables and to the pruning process (a point we will discuss in more detail elsewhere). LCM SW and LCM MLP deviate even more significantly, especially in regions where there have been very few, if any, observed transitions. Finally, CLUMondo and LCM LR follow the trend of the 'exact' curve. This is somewhat coincidental as these models are parametric, i.e., they perform a logistic regression for a specific type of curve, which is by chance similar to the one adopted here. Had we used a more complex probability distribution dependence on $\mathbf{y}$ instead, e.g., a bimodal distributions with two peaks, the result of LCM LR and CLUMondo would have been much less convincing (we checked this point).

Fig.~\ref{fig:cut_alloc} represents the transition probabilities estimated from the whole calibration-estimation/allocation comparison process. Let us point out that this is a single run (no average over a series of allocation for the same time step is performed), which by definition presents a significant statistical noise (section~\ref{sec:results_mae}) as this noise adds up at every stage of the whole process. This being said, once again, CLUMPY performs significantly better than the other modeling environments. We can observe very clearly the differences produced by the pruning parameter and the assumption of statistically independent explanatory variables for Dinamica EGO (we will quantify elsewhere the respective importance of these two sources of bias, which are usually the most significant). The $F=10$ (orange line) corresponds to the default value of this pruning parameter and produced a substantial bias in this example. We also notice that LCM LR and LCM MLP do not perform any allocation on this cut. Indeed, the LCM pruning method only selects the exact number of pixels to be transited (section~\ref{sec:pruning}). The pixels with the highest probability $\hat{P}(\mathbf{y}|u,v)$ being obviously not on this slice, no transition is observed, which is a very clear illustration of the bias involved in this allocation method.

\subsection{Execution Time Comparison}

Finally, we focus on the calculation times of the different models. Fast computations are important to be able to perform a sufficient number of simulations of the same problem (this is never done in LULC change modeling, but would in fact be required to extract meaningful statistical information, in agreement with the probabilistic and Markovian nature of projections). This would also allow the user to perform sensitivity analyses (something which again is never attempted).

\begin{table}
	\caption{Computation time of the different modeling environments for the case study of \ref{sec:case_study_def}, and for the calibration-estimation and global calibration/allocation comparison processes (total column) of section \ref{sec:evaluation_methods}. DE = Dinamica EGO with two different pruning factors ($F=10$ and $100$).}
	\label{tab:time}
	\centering
	\begin{tabular}{l|l|l}
		\multirow{2}{*}{model} & Calibration & \multirow{2}{*}{Total} \\
		& Estimation & \\
		\hline
		\hline
		\vspace{-0.3cm}& & \\
		CLUMondo & 45 sec & 45 sec \\
		\hline
		\vspace{-0.3cm}& & \\
		LCM MLP & 2 min 40 sec & 3 min 18 sec \\
		LCM SW & 19 min 15 sec & 19 min 53 sec \\
		LCM LR & 2 min 40 sec & 3 min 18 sec \\
		\hline
		\vspace{-0.3cm}& & \\
		DE F=10 & \multirow{2}{*}{32 sec} & 35 sec\\
		DE F=100 &  & 38 sec\\
		\hline
		\vspace{-0.3cm}& & \\
		CLUMPY & 6 sec & 11 sec\\
	\end{tabular}\\
\end{table}

The results are summarized in Table~\ref{tab:time}. First we point out that these results do not do justice to the major computational advantage of CLUMPY on very large problems (tens or hundreds of millions of pixels and several LULC transitions), where it outperforms all other modeling environments by a factor of at least 100 in computational time. However, for large problems, some of the methods evaluated do not even converge in 24h, while CLUMPY converges in a matter of minutes, which is why we chose a small enough problem, in order for the comparison to be possible. 

We still obtain a reasonable numerical efficiency for CLUMPY compared to the other models. Note that the KDE parameter $q$, which is set to $51$ here linearly influences the calibration-estimation time. Having a lower value of $q$ speeds up the process but increases the error in the estimated probability distribution. Also, the allocation algorithm presented in section~\ref{sec:methods_allocation} implies to  recompute $p(\mathbf{y}|u)$ frequently enough to obtain a bias-free allocation, and this also introduces a computation time penalty.

CLUMPY is always more efficient in computing dynamic distance maps than its closest competitor, Dinamica EGO. The largest the problem, the largest the gap in efficiency (CLUMPY is approximately quadratically more efficient than Dinamica EGO with increasing problem size). Conversely, the need to recompute probability distributions frequently enough during the allocation step (in order to ensure an unbiased allocation) is always a computation time penalty for CLUMPY. 
In fact, Dinamica EGO would greatly benefit from a change in its algorithm of dynamic distance updating (such as the python function \textsc{scipy.ndimage} used by CLUMPY).

\section{\uppercase{Conclusions}}
\label{sec:conclusion}

This paper introduces a new spatially explicit statistical LULC change modeling environment, CLUMPY. This environment is based on sound theoretical considerations, and is numerically efficient. In particular, we will show explicitly in a series of papers under preparation that our patch-oriented probabilistic formulation of LULC state transitions is formally correct, and that our algorithmic implementations of these theoretical bases is bias-free. We will also present an investigation strategy that allowed us to identify the sources of biases and errors in existing software in a systematic way, and correct them in our new modeling environment. This endeavor is designed to help reduce the differences of behavior between existing LULC change modeling environments on a given problem and set of data pointed out in the introduction, and to provide at least one such environment where remaining errors (mostly due to statistical noise) are under strict control and can be precisely quantified.

In the process, we have introduced a new calibration strategy inspired from the large body of work performed on density evaluation in the machine-learning community. Our implementation of this strategy produces significant improvements in the precision of the calibration process, in comparison to existing calibration methods. We have also used a new, bias-free, patch-allocation algorithm. This couple of calibration-allocation procedures is always unbiased and substantially more precise than existing ones. It is more efficient in terms of computational time on small (millions of pixels) problems, and significantly faster (up to $\sim 100$ times) than existing software on large or very large problems (tens to hundreds of millions of pixels).

We have finally proposed an evaluation method in section~\ref{sec:evaluation_methods} allowing us to perform effective comparisons of the performances of various modeling environments, including ours. This constitutes a first step towards a systematic validation procedure for LULC change models. This method takes advantage of the fact that it is both more relevant and more efficient to compare models in explanatory variable space rather than in physical space. Indeed, LULC change calibration data are often undersampled, by necessity, and the type of LULC change models analyzed here is statistical in nature. Both features imply that trying to reproduce transition locations exactly in physical space is often essentially impossible and misleading. Instead, one should focus on reproducing the correct probability structure in explanatory variable space, and, to a lesser extent, in patch parameter space (patch characteristics have not yet been seriously characterized in existing LULC change modeling environments). 

All these points will be elaborated upon in detail in our forthcoming papers. 

\bibliographystyle{apalike}
{\small \bibliography{biblio}}

\section*{\uppercase{Appendix}}
\vspace{-0.4cm}
\subsection*{Case Study Transition Probability Function}

The parameters used to define the function $P^*(v|u,\mathbf{y})$ in Eq.~\eqref{eq:f} are the following:
\begin{align}
\mu &= (150,0,0)\\
\Sigma &= \begin{pmatrix}
25^2 & 843 & 325 \\
843 & 10^2 & -16.3 \\
325 & -16.3 & 10^2 \\
\end{pmatrix}\\
\mathcal{D} &= \{\mathbf{y}~|~0 \le y_0 \leq 616,~0 \le y_1\leq 15,~0\le y_2\leq 60\}
\end{align}
\subsection*{Models Parameters}
\vspace{-0.4cm}
In section~\ref{sec:results}, we use various land use change models with the following parameters.

\subsubsection*{Dinamica~EGO}
\vspace{-0.4cm}
We have used version 5.2.1 and all the calculations were performed on a single CPU. The binning parameters are the following. The parameter \emph{increment} is fixed at $15$ meters for the elevation, $5^o$ for the slope and $10$ meters for the distance to urban areas. The \emph{minimum delta}, the \emph{maximum delta} and the \emph{tolerance angle} are respectively fixed at $50$, $500,000$ and $5.0$ for all explanatory variables.

\subsubsection*{Idrisi LCM}
\vspace{-0.4cm}
We have an Idrisi Selva license, which is relatively old (17.00). The estimation by logistic regression is done without sampling. The parameters of SimWeight are the default ones with notably the sample size fixed at $1000$. All the default parameters of MLP are kept.

\subsubsection*{CLUMondo}
\vspace{-0.4cm}
We have used version 1.4.0. The sampling parameter is fixed to $30\%$ of all observations. The \emph{number of cells distance between samples} is fixed to $2$ with no data values excluding and balanced sample enabled.

\subsubsection*{Clumpy}
\vspace{-0.4cm}
The KDE parameter $q$ (section~\ref{sec:kde}) is fixed to $51$. This is the only user-defined parameter in CLUMPY. This default value should be appropriate for most applications.

\end{document}